\documentclass[sigconf, nonacm]{acmart}

\newcommand\vldbdoi{XX.XX/XXX.XX}

\newcommand\vldbavailabilityurl{https://github.com/iDC-NEU/ATCC}
\newcommand\vldbpagestyle{plain}

\usepackage{enumerate}
\usepackage[linesnumbered, ruled, vlined, noend, resetcount]{algorithm2e} 
\usepackage{graphics}
\usepackage{balance}
\usepackage{graphicx}
\usepackage{subfig}
\usepackage{amsthm}
\usepackage{bm}
\usepackage{algpseudocode}
\usepackage{listings}
\usepackage{enumitem}
\usepackage{multirow}
\usepackage{float}
\usepackage{url}
\usepackage[normalem]{ulem}
\usepackage{comment}
\usepackage{amsmath}
\usepackage{epsfig}
\usepackage{booktabs}
\usepackage{lipsum}
\usepackage{array}
\usepackage{bbding} 
\usepackage{setspace}
\usepackage{makecell}
\usepackage{graphicx}
\usepackage{subcaption}

\newsavebox{\measurebox}

\newcommand{\Paragraph} [1] {\smallskip\noindent{\bf #1. }}

\newcommand{\zwx}[1]{\textcolor{black}{#1}}

\newcommand{\pzs}[1]{\textcolor{black}{#1}}

\newcommand{\oursys}{\texttt{ATCC}\xspace}

\begin{document}

\title{ATCC: Adaptive Concurrency Control for Unforeseen Agentic Transactions}


\author{Weixing Zhou, Zhiyou Wang, Zeshun Peng, Hetian Chen, Yanfeng Zhang, Ge Yu}
\affiliation{%
\institution{Northeastern University, China;}
}
\email{{zhouwx1, wangzy80, pengzs, chenht3}@mails.neu.edu.cn, {zhangyf, yuge}@mail.neu.edu.cn}

\begin{abstract}
Data agents, empowered by Large Language Models (LLMs), introduce a new paradigm in transaction processing. Unlike traditional applications with fixed patterns, data agents run online-generated workflows that repeatedly issue SQL statements, reason over intermediate results, and revise subsequent plans. To ensure data consistency, these SQL statements issued by an agent should be integrated into a transaction, referred to as \textbf{\emph{agentic transactions}}. Agentic transactions exhibit unforeseen characteristics, including long execution times, irregular execution intervals, and non-deterministic access patterns, breaking the assumptions underlying concurrency control (CC) (e.g., short-lived, predefined). Traditional CC schemes, which rely on fixed policies, fail to capture such dynamic behavior, resulting in inadequate performance.

This paper introduces \oursys, an adaptive \underline{C}oncurrency \underline{C}ontrol for \underline{A}gentic \underline{T}ransactions. \oursys continuously monitors and interprets the runtime behavior of each agentic transaction, evaluates its interactive phases, and dynamically adapts optimistic or pessimistic execution for each transaction. To ensure precise timing for adaptive switches, \oursys employs a reinforcement learning-based policy to balance immediate blocking against future abort costs. Additionally, to mitigate contention-induced tail latency and wasted reasoning cost caused by abort, a cost-aware priority-based lock scheduling is integrated to prioritize expensive or latency-sensitive transactions. Experimental results under agentic-like YCSB and TPC-C workloads demonstrate that ATCC improves the throughput of agentic transactions by up to four orders of magnitude and reduces tail latency by up to 90\% compared to state-of-the-art CC schemes.
\end{abstract}
%



\maketitle

\pagestyle{\vldbpagestyle}



\section{Introduction}
\label{section1}
\begin{figure}[t]
\centering
\includegraphics[width=0.45\textwidth]{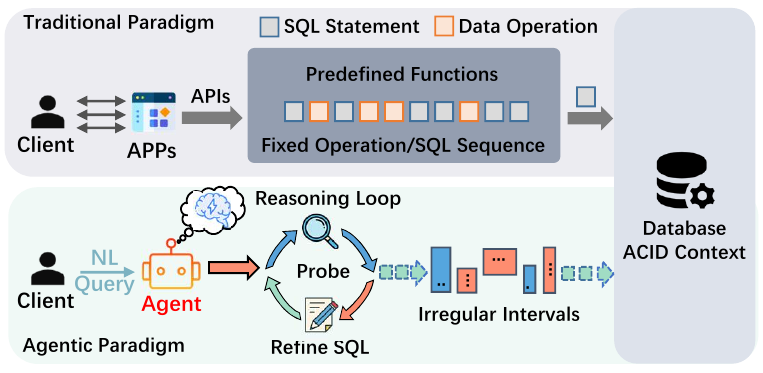}
\caption{Comparison of traditional predefined transaction processing (TP) and agentic TP with dynamic SQL generation.}
\label{fig:interactive-cmp}
\end{figure}

Powered by advances in Large Language Models (LLMs) \cite{brown2020language, zhao2023survey, radford2019language, touvron2023llama}, the emergence of data agents \cite{yao2023react, yao2023tree, schick2023toolformer} represents a fundamental shift in the paradigm of data management. These agents, equipped with advanced perception, reasoning, and learning capabilities, can interpret high-level user goals and compile them into precise executable workflows (e.g., text2SQL \cite{liu2025skyrlsql, li2025continuumefficientrobustmultiturn, sun2025rethinkingtexttosqldynamicmultiturn, wang2021tracking, zhong2017seq2sql, li2025omnisql, gao2023text, li2024dawn, guo2019towards, liu2024survey}.
As illustrated in Figure~\ref{fig:interactive-cmp}, traditional applications issue a fixed set of SQL statements through predefined functions. In contrast, given a high-level intent, an agent can autonomously plan and execute tasks. By examining the intermediate results, the agent iteratively revises its plan until the task is completed. This agent-driven execution reduces user effort and enables more complex, personalized workflows by adapting task generation and decision logic to user preferences and context.

Agents run online-generated workflows that iteratively generate SQL statements and dynamically adjust their decisions based on real-time feedback. In high-stakes domains such as finance, e-commerce settlement, and supply chains, ensuring a consistent data view and atomic state transitions is crucial during agent task execution. Without transactional guarantees, the agent may read inconsistent snapshots or lost partial updates, leading to incorrect decisions. Consequently, the data operations issued by an agent to complete a task should be encapsulated as a single transaction to ensure correctness. This emerging class of agent-driven transactions, referred to as \textbf{\textit{agentic transactions}}, introduces a new type of transactional behavior.

\textbf{An example.} Consider a digital banking scenario where an agent adjusts the credit limit of a corporate customer. The agent first queries historical records to build the customer's profile, then retrieves the customer's latest financial status (e.g., recent payment behavior, market conditions), and finally updates the credit limit. Without ACID guarantees, the agent may approve a credit adjustment based on stale or incorrect status. This can lead to incorrect limit assignments, compliance violations, and financial risks.

Agentic transactions execute in a sequence of phases, leading to execution characteristics that fundamentally differ from traditional transactions. While traditional OLTP transactions feature interleaved read/write operations, agentic transactions exhibit a distinct \emph{phase structure} with shifting read/write intent: (i) a read-intensive \emph{explore} phase to gather context, (ii) a \emph{refine} phase to narrow down candidate plans, and (iii) a write-intensive \emph{commit} phase to apply final updates. Consequently, this phased execution model may lead to (i) \textit{\textbf{long execution times}}, (ii) \textit{\textbf{irregular execution intervals}}, and (iii) \textit{\textbf{non-deterministic access patterns}}.
During the refine phase, agents perform multiple rounds of reasoning to refine their execution plans and dynamically adjust SQL statements to better align with user intent. This causes SQL structures and execution order to change dynamically. Since the duration and number of reasoning rounds depend on the accuracy of intermediate results, the timing of SQL statement execution is inherently irregular. Consequently, this iterative reasoning process may result in extended transaction lifetimes. A detailed analysis is presented in Section~\ref{agentic_transaction}.

However, these characteristics break the static assumption (e.g., predefined access patterns, short-lived transactions) of the underlying traditional concurrency control (CC). Consequently, existing CC schemes adopted by mainstream databases fail to handle agentic workloads efficiently:
(i) Pessimistic Concurrency Control (PCC) \cite{yu2018sundial, lu2018star, ren2012lightweightlocking, bernstein1987concurrency} prevents conflicts via locking but leads to long blocking and inflated tail latency.
(ii) Optimistic Concurrency Control (OCC) \cite{silo, yu2016tictoc, Hekaton, Cicada, Polaris} avoids blocking but incurs high abort rates under frequent conflicts and stale reads.
(iii) Hybrid schemes \cite{Polyjuice, ComposingCC, combiningCC, CormCC, Plor, mocc} attempt to apply PCC/OCC to hot and cold data partitions, or mix optimistic reads with pessimistic writes to achieve efficient transaction execution. However, the irregular execution intervals and evolving access patterns cause contention windows and hotspots to shift during execution.

Therefore, agentic CC must go beyond static OCC/PCC and instead \textbf{observe and react to runtime behavior}.  

In this paper, we propose \oursys, an adaptive concurrency control algorithm that dynamically applies optimistic or pessimistic approaches for transactions.
\oursys continuously estimates transaction states (e.g., execution phases, conflict and abort risk) from transaction runtime features and the current system state, and selectively acquires locks on data items, thereby reducing blocks and aborts of agentic transactions with minimal impact on system performance.

In \oursys, accurately assessing transaction states and locking timing is crucial. Premature locking causes long blocking, while delayed locking increases stale reads and abort risk. For agentic transactions, the unforeseen characteristics make fixed heuristics infeasible. 
Learning-based approaches \cite{QueryForecasting, Scheduling2019, AI4DB}, which can associate observed execution context with downstream outcomes and adapt from runtime feedback under dynamic and stochastic behavior, are better suitable for agentic transactions.
Thus, we use a reinforcement learning (RL) model to \emph{learn} locking decisions, capturing the delayed trade-off between immediate blocking and future abort cost. 
By leveraging the distinctive phase structure of agentic transactions, \oursys learns a phase-aware adaptive locking policy that appropriately employs locks, balancing optimistic execution with targeted locking to reduce aborts without excessive blocking.

Even with adaptive locking, agentic transactions still face blocking or aborting due to lock contention. Blocking long-running or repeatedly re-executed transactions exacerbates tail latency, while aborting transactions that have already incurred multiple reasoning and extensive data operations wastes substantial computation resources and increases user costs.
To address this, \oursys integrates priority-based locking to guide lock scheduling under contention. Specifically, \oursys derives a transaction priority by jointly considering runtime factors such as execution time and estimated reasoning cost, and uses this priority to optimize lock scheduling. This approach prioritizes critical transactions, accelerates their processing, and reduces tail latency.
In summary:

\begin{itemize}[leftmargin=*]
\item We identify the specific characteristics of agentic transactions and reveal their distinct phase structure, enabling adaptive CC. 
\item We present \oursys, an adaptive concurrency control mechanism that integrates with an RL-based policy that learns phase-aware locking timing and a priority-based locking scheme to enable efficient execution for agentic transactions.
\item We implement \oursys on openGauss \cite{openGauss-MOT} and compare it with SoTA methods on agentic workloads. Experiments show \oursys improves the performance of agentic transactions up to 4 orders of magnitude higher throughput and 90\% lower tail latency. 
\end{itemize}
\section{MOTIVATION}
\label{section:2}

\begin{figure}[t]
\centering
\includegraphics[width=0.45\textwidth]{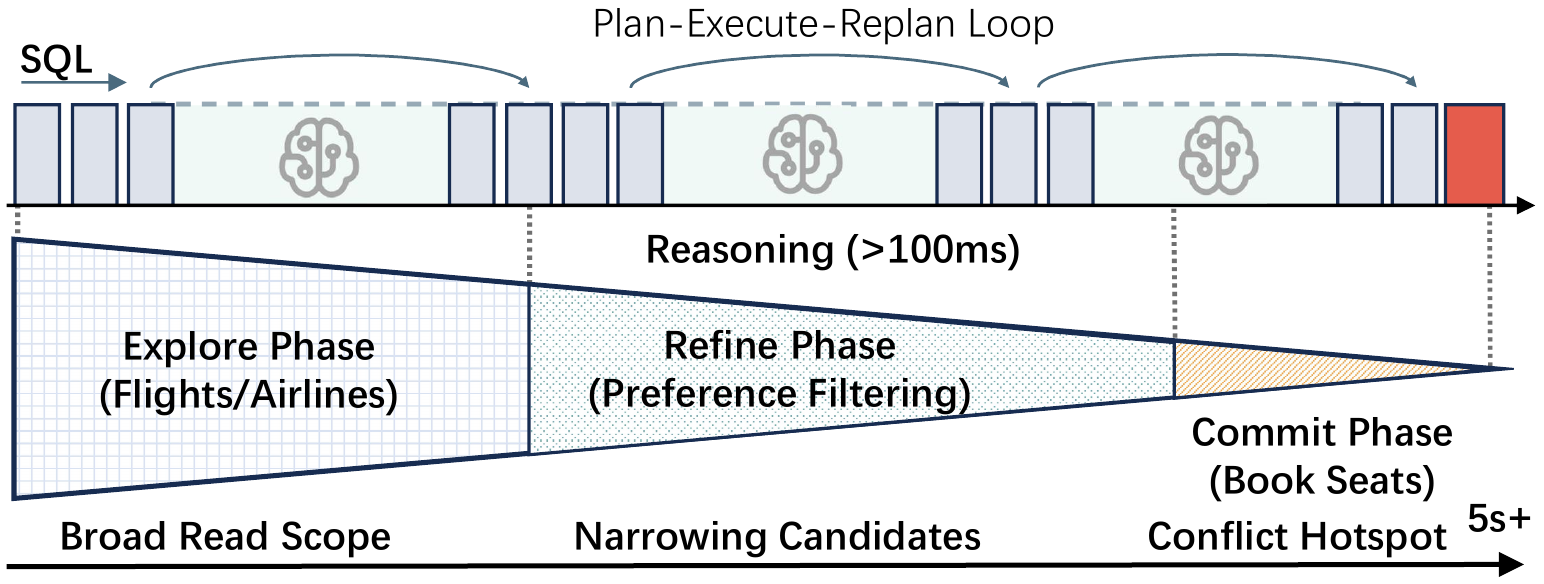}
\caption{An execution illustration of agentic transactions.}
\label{fig:interactive-phase}
\end{figure}

LLM-based agents offer an effective way to offload complex decision-making from end users to automated workflows. 
To ensure data consistency, the agent workflow should be encapsulated as a single transaction, forming an \textbf{\emph{agentic transaction}}. However, because SQL statements are generated during LLM reasoning, agentic transactions are inherently unpredictable and violate the static assumptions (e.g., definitions and predictable access patterns) of traditional database workloads.

To characterize the features of agentic transactions, we implement an agentic flight-booking workload. Starting from an intent (e.g., departure city, destination, budget, or time constraints), the agent autonomously explores candidates, refines constraints, and finalizes a reservation (see Section \ref{section:Evaluation} for details on workloads).

\subsection{Features of Agentic Transactions}
\label{agentic_transaction}

Agentic transactions exhibit several distinctive characteristics that differ from traditional OLTP workloads:

\textbf{Long and highly variable lifetimes.}
As shown in Figure \ref{fig:interactive-phase}, the agent interleaves SQL execution with multiple reasoning and replanning steps. The additional reasoning steps extend end-to-end lifetimes by $4$--$5$ orders of magnitude compared ($\sim 20$ s in average) to backend transactions ($\sim$ 0.3 ms).

\begin{figure}[t]
\centering
\includegraphics[width=0.47\textwidth]{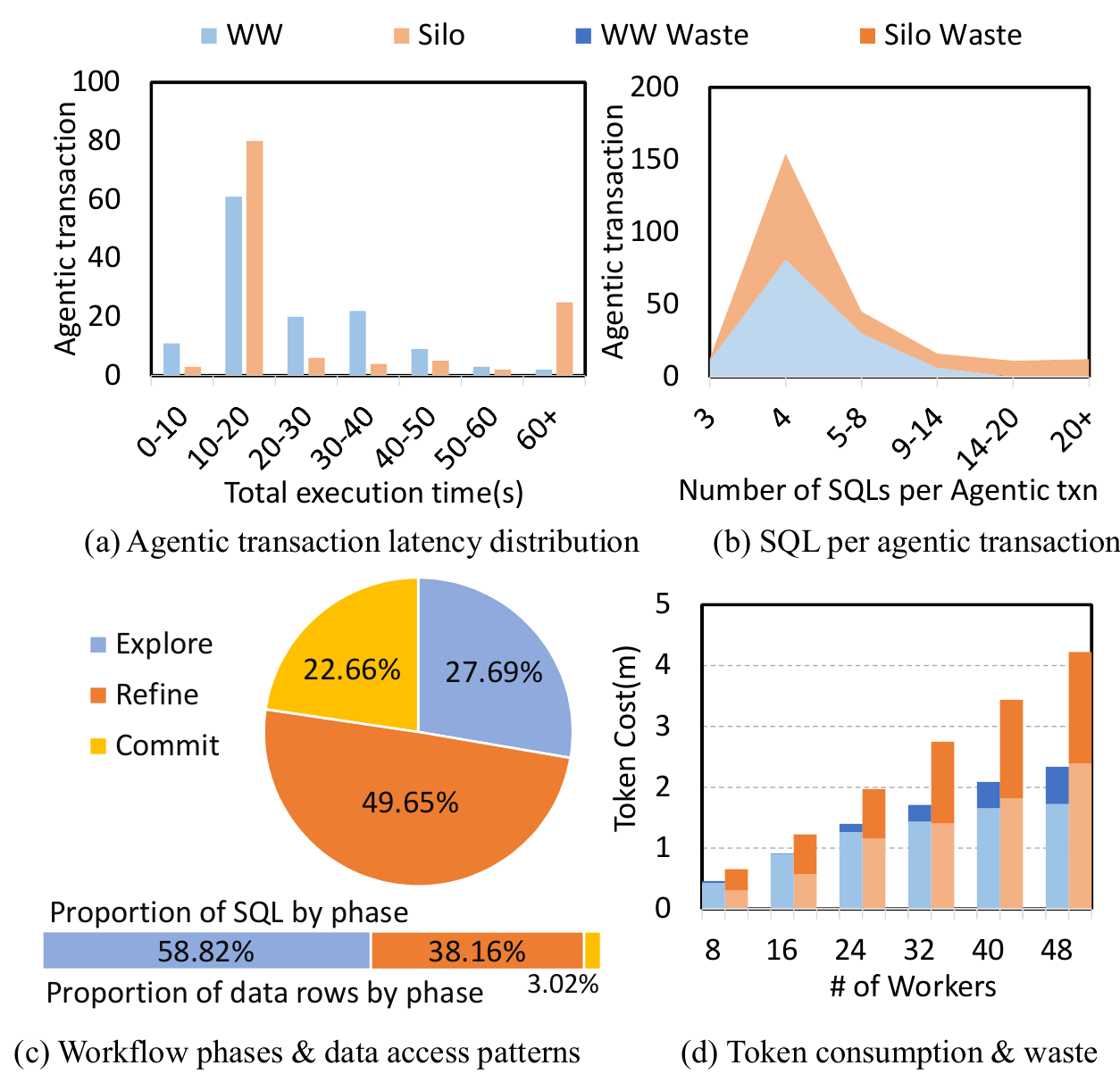}
\caption{Experimental results of an agentic flight-booking workload.}
\vspace{-0.1in}
\label{fig:agentic_analyze}
\end{figure}

\textbf{Irregular intervals.}
Unlike traditional OLTP transactions that execute SQL statements within milliseconds, the execution intervals of agentic transactions are dominated by reasoning time and additional agent-side data operations, which vary significantly. 
This leads to the execution time spans a wide range for OCC, from $10$~s to $60$~s in Figure \ref{fig:agentic_analyze}a.

\textbf{Non-deterministic and evolving access patterns.}
Data agents refine the execution plans based on intermediate results, issuing $4$-$20$ SQLs compared to $3$ for backend transactions (Figure \ref{fig:agentic_analyze}b). Moreover, even under the same user intent, the plan overlap across runs is below 50\%, with significant variability in query sequences. This divergence is driven by environment dynamics (e.g., changing flight availability/prices) and LLM-induced variability in query construction and candidate selection.

\textbf{Phase structure with shifting read/write.}
In the early \emph{explore} phase, the agent performs coarse, scan-intensive reads with broad predicates targeting metadata that is rarely updated, accounting for about $49\%$ of all SQLs and $58\%$ of all accessed row data items. In the mid \emph{refine} phase, agents tighten predicates to retrieve accurate details and generate a refined shortlist, contributing $27\%$ of SQLs and $38\%$ of rows. In the \emph{commit phase}, the agent executes booking logic, with $22\%$ of queries but over $90\%$ of writes, accessing just $3\%$ of rows.  This phased read then write pattern contrasts sharply with traditional OLTP transactions, where reads and writes are typically interleaved within short, predefined workflows.

\textbf{Substantially high abort costs.}
Aborting traditional transactions typically wastes only a few milliseconds of execution time. In contrast, aborting agentic transactions invalidates the entire multi-round reasoning process. In our evaluation, recovering from a failure often requires re-executing more than $8-15$ SQL statements and, in the worst cases, wasting up to $100.3K$ LLM tokens. As a result, the per-abort cost for agentic transactions is much higher. As shown in Figure \ref{fig:agentic_analyze}d, the frequent aborts and retries under OCC (Silo) lead to substantially higher token costs.

\subsection{Challenge and Opportunity}
\label{Challenges-and-Opportunities}
Agentic transactions, with unforeseen characteristics and high retry cost, pose significant challenges for CC schemes. The system must make appropriate CC decisions without prior knowledge of the complete access set to reduce aborts. A natural approach is to learn CC decisions via reinforcement learning (RL), as CC is an online sequential decision problem with delayed consequences. 
By learning per-operation execution states (e.g., the sequence of executed operations, inter-transaction dependencies) from the performed optimistic/pessimistic actions and the eventual commit/abort outcomes, RL can learn the trade-off between immediate blocking and future aborts under uncertainty. 

However, such operation-granularity RL is effective when access patterns are stable (e.g., predefined workloads). Under agentic workloads, the non-deterministic access patterns lead to an unbounded state space and poor generalization.
Moreover, agentic transactions are not entirely random. Our measurements reveal a distinct \textbf{\emph{phase structure}}. This offers a key opportunity: the system can dynamically adapt CC strategies based on the inferred phase and the estimated conflict or abort costs.
For example, the read-heavy exploration phase benefits from lightweight, non-blocking execution; the refinement phase, focused on protecting reads, ensures data consistency; and the write-intensive commit phase favors comprehensive locking to prevent costly aborts.

Building on this insight, \oursys models CC as a phase-aware sequential decision-making problem. It uses RL to infer the current execution phase, contention level, and abort cost to decide when and what to lock. By dynamically adapting between optimistic and pessimistic execution, \oursys effectively reduces execution costs and is well-suited to the dynamic, evolving nature of agentic workloads.

\section{Overview}
\label{section:3}
\label{Overview}

\begin{figure}[t]
    \centerline{\includegraphics[width=0.43\textwidth]{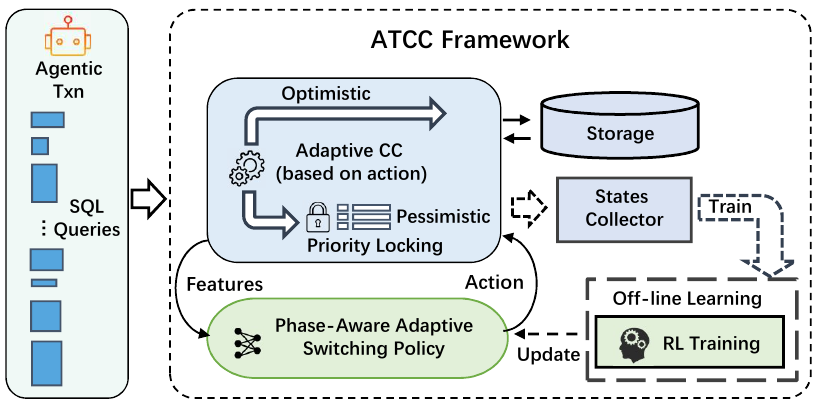}}
    \caption{The overall architecture of ATCC.}
    \Description[]{}
    \label{fig-overview}
\end{figure}

In this section, we provide the design and overall architecture of \oursys.
As illustrated in Figure \ref{fig-overview}, \oursys applies an adaptive concurrency control framework that dynamically selects OCC or PCC for data operations. 
As discussed in Section \ref{Challenges-and-Opportunities}, \oursys learns a phase-aware adaptive policy to evaluate the transaction's current state (e.g., execution phases, the cost of abort) and obtains guidance actions.
Based on the inferred actions, \oursys adjusts the transaction’s execution strategy, including: (i) executing reads/writes optimistically or pessimistically based on a selected locking range; and (ii) boosting the transaction’s priority to enable lock preempting with Wound-Wait Locking.
Additionally, to better accommodate the dynamic nature of agentic transactions and optimize system performance, \oursys periodically gathers execution statistics and re-trains offline to refine the adaptation policy.

\Paragraph{\textbf{Phase-aware adaptive policy}}
Agentic transactions exhibit multi-round execution, during which their access patterns and conflict risk evolve over time. With RL models, \oursys adaptively blends optimistic and pessimistic execution over the transaction lifetime to jointly optimize interactivity, throughput, and retry cost. Unlike operation-level RL policies that condition decisions on per-access identifiers that face an unbounded state space, \oursys leverages phase structure as the key generalizable signal under non-deterministic access patterns, shifts the RL problem from operation prediction to phase inference estimation. With a phase-aware state representation (Section \ref{state_space}), the RL state space remains compact and consistent across diverse transaction access patterns, enabling the learned policy to be more efficient under agentic workloads.

\Paragraph{\textbf{Contention-oriented selective locking}}
With RL, \oursys can analyze the execution phases of transactions and adaptively apply locks to avoid stale reads and conflicting concurrent writes.
The locking actions implemented for different stages of a transaction will affect system performance. 
When a transaction enters a critical phase and directly locks all access data without considering the current contention level, this leads to high lock management overhead and unnecessary blocking.
Further, such a locking strategy lacks the flexibility to adapt to varying contention levels and execution phases.
To balance learnability and effectiveness, \oursys employs contention-oriented selective locking, dynamically adjusting lock granularity based on the inferred phase and current contention level (Section \ref{action_space}).

\Paragraph{\textbf{Abort-cost-aware priority scheduling with deferred updates}}
Lock-based execution can introduce long blocking, deadlocks. In such cases, randomly aborting a transaction with longer execution times, multiple reasoning steps, or extensive read/write sets tends to incur higher abort and retry costs. To account for this, \oursys integrates an abort-cost-aware, multi-level priority scheduling into the Wound-Wait locking scheme, to prioritize critical transactions (Section \ref{action_space}).
Another consideration is that transactions may hold write locks for long periods due to reasoning or lock blocking. This will block backend readers. To mitigate this, \oursys uses deferred updates to enable the latest committed version reading, achieving fast execution for backend transactions (Section \ref{delayed_updates}).

\section{Phase-Aware Adaptive Switching Model}
\pzs{Reinforcement learning (RL) learns a control policy through trial-and-error interaction with the environment to maximize numerical rewards. Its core components include: states (observations of the environment), actions (decisions the RL agent can take), a policy (mapping observed environmental states to actions), a reward signal (defining the optimization objective), and an environment (with which the RL agent interacts). At runtime, the RL agent observes a state, takes an action, receives a reward, and transitions to a new state. When applying RL, the design of the state and action spaces, and the reward function, critically determines the effectiveness and robustness of the learned policy.}

\pzs{In the following, we present the design of these components.}


\subsection{\pzs{Phase-Aware State Representation}}
\label{state_space}
\pzs{A key challenge in designing an RL-based CC is capturing the characteristics of agentic workloads. Different agentic workloads exhibit distinct access patterns, and even within the same transaction, execution phases vary significantly in abort costs and access contention. To enable the policy to make phase-specific CC decisions, we design a state representation that captures (i) phase-related signals (e.g., query intervals) and (ii) contention indicators (e.g., retry count) to apply optimistic or pessimistic execution adaptively.}

\Paragraph{Phase-related signals}
\pzs{The following signals are designed to infer the current execution phase of an agentic transaction.
\begin{itemize}[leftmargin=*]
\item \textit{Query interval time.} The elapsed time between the completion of the previous SQL and the start of the current one, which is used to estimate abort cost. Longer intervals indicate more LLM reasoning time, and thus higher retry costs.
\item \textit{Read/write-set size delta.} The increment of the read/write set size per SQL round. In the explore phase, the read set expands rapidly. In the refine phase, the read-set and write-set sizes are relatively stable. In the commit phase, the write set expands rapidly.
\item \textit{Read/write-set overlap ratio.} The overlap ratio of accessed rows between consecutive SQL rounds. A low overlap ratio indicates the transaction is exploring new data, typically in the explore phase. A high overlap ratio indicates the transaction is converging on a stable result set, typically in the refine or commit phase.
\item \textit{Consecutive writes.} A signal detects whether the transaction has issued multiple writes. Indicates whether the transaction has transitioned from the refine phase to the commit phase.
\end{itemize}}



\Paragraph{Contention indicators}
\pzs{The following signals are designed to evaluate the transaction's contention level and abort probability.
\begin{itemize}[leftmargin=*]
\item \textit{Hotspot access ratio.} The proportion of hotspot records in the read/write sets. Transactions are more likely to be aborted due to conflicts when accessing more hotspot records.
\item \textit{Blocking time.} The total time a transaction spends waiting for locks. A longer blocking time indicates higher data contention.
\item \textit{Retry Count and Signal.} The total number of times a transaction has restarted due to conflicts (e.g., OCC validation or PCC lock failures). Repeated failures indicate high contention and require a higher CC priority to reduce the risk of abort.
\item \textit{Global metrics.} Database metrics including the abort rate, throughput, average and tail latency, and lock-queue length. These metrics provide a global view of workload contention.
\end{itemize}
}



\begin{table}[t]
\centering
\caption{Contention-driven actions in \oursys}
\small
\label{tab-actions}
\begin{tabular}{ccc}
\hline
\textbf{Action} & \multicolumn{2}{c}{\textbf{Description and Rationale}} \\ 
\hline
\textbf{Remain in OCC Mode} & \multicolumn{2}{c}{Continue under OCC} \\
\multirow{2}{*}{\textbf{Locking Strategies}} & {Hot Read set} & {Cold Read set} \\ 
{} & {Hot Write set} & {Cold Write set} \\
\hline
\textbf{Prioritize Transaction} &  \multicolumn{2}{c}{Boost a transaction's priority.} \\ 
\hline
\end{tabular}
\end{table}

\subsection{Contention-Aware Switching Policy}
\label{action_space}
Table~\ref{tab-actions} shows the action space.
\pzs{To achieve high performance when executing agentic transactions, the action space enables elastic CC switching from optimistic to a pessimistic mode based on phase-related signals. OCC enables high concurrency by allowing transactions to execute without acquiring locks. All transactions initially start in OCC mode. This is because transactions are predominantly read-only in the explore phase. We can utilize deferred writes (see Section \ref{delayed_updates}) to enable lock-free reads, avoiding locking overhead.}

\pzs{As transactions progress through multiple SQL rounds and transition into the refine and commit phases, the accumulated execution cost makes transaction aborts increasingly expensive. Therefore, when transactions are likely to conflict (e.g., modifying hotspot records), they are switched to a more pessimistic mode by acquiring priority locks on accessed records, improving commit rates.}


\Paragraph{Lock granularity}
\pzs{To increase concurrency, pessimistic transactions do not lock all accessed rows. Instead, lock granularity is dynamically adjusted based on workload patterns (read/write intensity and hotspot frequency). \oursys can identify hotspot read/write tuples (see Section~\ref{metadata}) and apply different locking strategies accordingly. For example, under high contention, transactions can lock their entire write sets and hot read sets to reduce abort rates, while cold data remains unlocked to preserve concurrency.}

\Paragraph{Lock priority}
\pzs{\oursys employs priority-based pessimistic locks to balance concurrency with the commit rate of critical transactions. To prevent starvation from lock blocking, we adopt the Wound-Wait locking strategy. Specifically, we maintain a priority score for each transaction $T$ to guide lock acquisition decisions.}

{\footnotesize
\begin{equation}
\label{eq:priority}
\mathrm{Prio}(T)\!=\!
\lfloor \alpha \!\cdot\! \frac{\mathrm{SQL}(T)}{\Delta_s} \rfloor
\!+\! \lfloor \beta \!\cdot\! \frac{\mathrm{Blocked}(T)}{\Delta_b} \rfloor
\!+\! \lfloor \lambda \!\cdot\! \mathrm{Retry}(T) \rfloor
\!+\! \lfloor \rho \!\cdot\! \frac{\mathrm{Interval}(T)}{\Delta_i} \rfloor
\end{equation}
}

\zwx{As shown in Equation~\ref{eq:priority}, $\mathrm{SQL}(T)$ measures the accumulated database operation cost, reflecting work intensity; $\mathrm{Blocked}(T)$ tracks the cumulative blocking time; and $\mathrm{Retry}(T)$ counts abort-restart times. These metrics identify potential starvation and elevate priority to reduce tail latency. $\mathrm{Interval}(T)$ estimates the LLM reasoning time between SQL rounds. In our setting, we estimate token consumption from reasoning time, where longer reasoning times indicate higher abort costs. The parameters $\Delta_s, \Delta_b, \Delta_i$ are time quanta used to discretize each signal.} 


\subsection{Cost-Aware Reward Design}
\label{reward}
\pzs{The reward function must account for abort costs. Unlike traditional workloads, where transaction aborts only waste database resources, aborting an agentic transaction discards both the database work and the expensive LLM reasoning. These costs include not only traditional workload metrics (e.g., throughput, latency) but also LLM reasoning overhead. \oursys incorporates these costs into the reward function to accurately reflect the actual execution cost.}


\pzs{At each decision step $t$, the reward $r_t$ is defined in Equation~\ref{eq:reward}.}
{\footnotesize
\begin{equation}
\label{eq:reward}
\begin{aligned}
r =
&(\beta_1  \cdot \mathbf{1}_{\text{commit}} - \beta_2 \cdot \mathbf{1}_{\text{abort}})
- \alpha_1 \cdot \Delta \text{Latency}_t
+ \alpha_2 \cdot \Delta \text{TPS}_t
\\
&
- \alpha_3 \cdot \Delta \text{AbortRate}_t
+ \alpha_1 \cdot \frac{\mathrm{SQL}(T)}{\Delta_s} 
+ \alpha_2 \cdot \frac{\mathrm{Interval}(T)}{\Delta_i} 
\end{aligned}
\end{equation}
}

\pzs{We explicitly model retry costs to determine whether to block (using locks) or abort a transaction. Specifically, we use $\mathrm{SQL}(T)$ to measure the accumulated database execution cost and $\mathrm{Interval}(T)$ to estimate the LLM reasoning overhead between SQL rounds. Traditional transaction metrics (\(\mathbf{1}_{\text{commit}}\), \(\mathbf{1}_{\text{abort}}\)) and database metrics (i.e., \(\Delta \text{TPS}_t\), \(\Delta \text{Latency}_t\), and \(\Delta \text{AbortRate}_t\)) are incorporated into the optimization objective to reflect the efficiency of global scheduling.}




\subsection{Low-Latency Model Invocation}
\label{model_invoking}
Invoking the RL model takes several milliseconds. Because RL model invocation lies on the critical path during transaction execution, it significantly degrades transaction throughput. Moreover, massive model invocations consume substantial resources, further impacting transaction performance.

To reduce model invocation overhead, \oursys proposes a two-layer invocation mechanism. First, \oursys applies a state discretization strategy: by learning phase boundaries, the system maps high-dimensional, potentially unbounded runtime features into a finite set of discrete states and compiles the learned policy into a compact lookup table. After completing a batch of operations, transactions combine their features into state keys and fetch actions through quick table lookups. 
Second, \oursys integrates an asynchronous model with real-time workload awareness to refine policy decisions. After each transaction execution, the system asynchronously reports transaction characteristics to the model, which corrects the action obtained from the table lookup.
This decouples model inference from transaction execution through asynchronous reasoning, removing it from the critical path.

\section{Adaptive Concurrency Control}

\label{execution}
\begin{figure}[t]
\centerline{\includegraphics[width=0.39\textwidth]{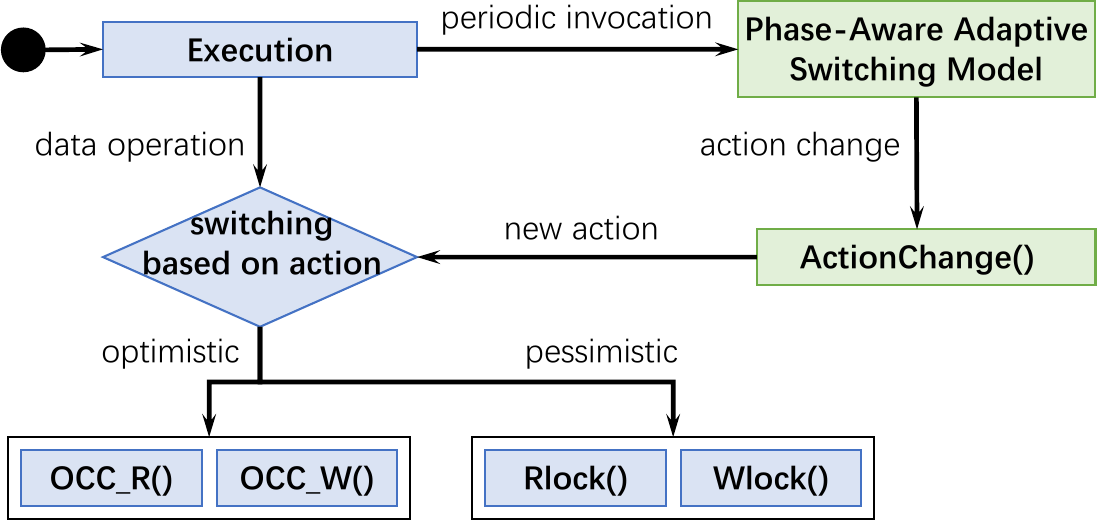}}
\caption{The execution flow and function calls of adaptive concurrency control.}
\Description[]{}
\label{fig-execution}
\end{figure}

\subsection{Metadata for Fine-Grained Adaptation}
\label{metadata}
Our system maintains DRAM metadata for each transaction and data item (i.e., a record) to enable adaptive execution.

\oursys launches a worker thread for each transaction received from a client. Each worker thread maintains a context (ctx) that includes the worker ID ($wid$, 16-bit), the start timestamp ($startTS$, 47-bit), and the transaction state ($status$, 1-bit) to indicate whether the transaction is running or aborted (e.g., toggled under WOUND\_WAIT for conflict resolution). 
The $wid$, $startTS$, and $status$ together form a 64-bit integer (transaction ID, $tid$), which uniquely identifies the current running transaction.
The context also tracks state space variables, including the transaction priority ($priority$) and the current action ($action$). The contexts of all worker threads are stored in an array indexed by worker ID, which is globally accessible.
The transaction execution results are stored in the read/write sets ($RS$ and $WS$). In addition, we maintain the hot read/write sets ($RHS$ and $WHS$) to perform action changes efficiently.

\oursys also maintains DRAM metadata for each data item, including a data version, a read lock counter ($R\_count$), and a write lock owner ($W\_owner$) that track the transactions holding read and write locks. 
A read list ($R\_list$) is maintained to record active read transactions, while a lock queue ($lock\_queue$) manages transactions waiting to acquire pessimistic locks.
To manage optimistic validation and pessimistic writes correctly, \oursys relocates the OCC latch bit (the committing latch) from the data version (in the traditional OCC prototype) to the write lock. When a transaction successfully acquires the write lock and enters the commit phase, it sets this latch bit to \texttt{true}, entering a non-preemptive committing state that prevents lock preemption under the Wound–Wait policy (details are discussed in Section \ref{validation}).
To adaptively manage hotspots, \oursys uses Hot Flags ($hot\_flag$) to categorize data items as cold or hot and periodically evaluate access patterns to update the hotspot status.

\subsection{Hybrid Execution with Deferred Writes}
\label{delayed_updates}
As shown in Figure \ref{fig-execution}, \oursys adaptively performs optimistic or pessimistic operations based on the rules specified by the action. 
In general, transactions are executed optimistically (init action) and periodically invoke the RL model.
When a transaction's action changes, the worker validates the data items that need to be locked to detect existing conflicts, and then acquires the corresponding locks (Section \ref{switch_validation}). Upon committing, the worker validates the remaining portions of the data items (those that are unlocked) and completes the final commit process (Section \ref{validation}).

\begin{figure}[t]
\centerline{\includegraphics[width=0.43\textwidth]{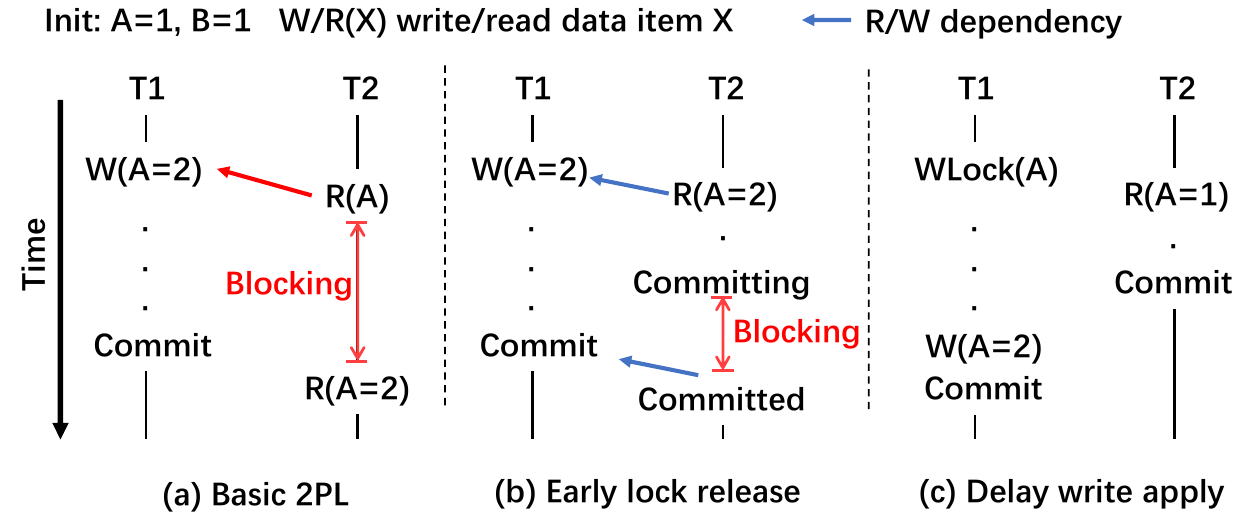}}
\vspace{-0.1in}
\caption{Comparison of transaction execution under different locking mechanisms: (a) basic 2PL, (b) early lock release, and (c) delayed write apply.}
\Description[]{}
\label{fig-locking}
\vspace{-0.1in}
\end{figure}

\Paragraph{Non-Blocking Optimistic Execution via Deferred Writes}
In \oursys, optimistic reading and pessimistic locking coexist. For long-running agentic transactions, traditional pessimistic locking can easily lead to severe blocking. As shown in Figure \ref{fig-locking}a, transaction $T_1$ and $T_2$ operate on the same data item $A$, where $T_1$ writes $A$ and $T_2$ reads $A$. Once $T_1$ acquires the write lock on $A$ and updates it, $T_2$ must wait until $T_1$ releases the lock. For agentic transactions with a prolonged lifetime, holding locks for an extended period leads to head-of-line blocking and poor responsiveness for other transactions.

Early lock release \cite{PWV, Bamboo} (Figure~\ref{fig-locking}b) aims to reduce blocking by releasing locks before a transaction commits. After $T_1$ completes its updates, it releases the lock on $A$, allowing $T_2$ to read the new value without waiting for $T_1$ to commit its changes. However, to maintain correctness, $T_2$ is not permitted to commit before $T_1$ does. Otherwise, if $T_1$ aborts, $T_2$ would encounter a dirty read, violating isolation guarantees. In agentic workloads, this design also introduces long blocking, as $T_2$ may finish its own logic early but still must wait for the long-running transaction $T_1$ to commit.

To avoid long blocking, \oursys adopts a \emph{delayed write apply} mechanism (Figure~\ref{fig-locking}c). Transactions acquire write locks to resolve concurrent write-write conflicts, but buffer the new value in the local write set instead of immediately updating the database. Short-lived transactions like $T_2$ can read the last committed value ($A = 1$) without being blocked, achieving faster commit. 

To this end, optimistic reads are processed directly without checking whether a transaction owns the write lock, but transactions still need to spin on the committing latch to check if a writer is currently applying an update, thereby avoiding reads of in-flight, partially updated versions. For optimistic writes, the transaction directly inserts the row into the write set.


\begin{algorithm}[t]
\begin{spacing}{1}
\SetAlgoNoLine
\small
    \caption{Functions of Pessimistic Locking in \oursys}
    \Description[]{}
    \label{alg-TransProcess}
    \KwIn{thread (transaction) context $ctx$}
    \KwOut{return true if success; false otherwise}

    \SetKwFunction{FRLock}{RLock}
    \SetKwFunction{FWLock}{WLock}



    \SetKwProg{Fn}{Function}{:}{\KwRet}
    \Fn{\FRLock{Row $row$}}{
        \If{$W\_owner$ == $tid$ or $tid$ $\in$ $reader\_list$}{
          \Return; \textcolor{gray}{ //already get the lock.} \\
        }
        \If{$W\_owner$ $\neq$ INVALID}{
            $writer$ = ctx\_arr[$W\_owner.wid$]\;
            \If{$writer.priority$ < $ctx.priority$}{
                Abort($writer$)\;
            }
            \uElse {
                Insert($lock\_queue$, $ctx$, Read) and Wait\; 
                \textcolor{gray}{ //wait until get the lock or abort by other txn.}\\
            }
        }
        $reader\_list$.add($ctx.tid$)\;
        $R\_count$ += 1\;
        Insert($row$, $ctx.RS/HRS$)\;
    }

    \SetKwProg{Fn}{Function}{:}{\KwRet}
    \Fn{\FWLock{Row $row$}}{
        \If{$W\_owner$ == $tid$}{
          \Return\;
        }
        \If{$R\_count$ $\neq$ 0}{
            \ForEach{$R\_tid$ in $reader\_list$}{
                $reader$ = ctx\_arr[$R\_tid.wid$]\;
                \If{$reader.priority$ > $ctx.priority$}{
                    Insert($lock\_queue$, $ctx$, Write) and Wait\;
                }
            }
            \ForEach{$reader$ in $reader\_list$}{
                Abort($reader$)\;
                $R\_count$ -= 1\;
            }
        }
        \If{$W\_owner$ is INVALID}{
            $W\_owner$ = $tid$\;
        }
        \uElse {
            $writer$ = ctx\_arr[$W\_owner.wid$]\;
            \If{$writer.priority$ < $ctx.priority$}{
                Abort($writer$)\;
                $W\_owner$ = $tid$\;
            }
            \uElse {
                Insert($lock\_queue$, $ctx$, Write) and Wait\;
            }
        }
        Insert($row$, $ctx.WS/HWS$)\;
    }
    
\end{spacing}
\end{algorithm}

\Paragraph{Priority-Based Preemptive Locking}
In \oursys, the priority-based locking is used to resolve conflicts among pessimistic operations.

For read operations, the delayed write-apply mechanism enables short-lived transactions to read optimistically, facilitating fast execution. However, agentic transactions still acquire explicit read locks on critical items to avoid aborts during their long execution, where other short or lower-priority transactions might update these items before the agentic transactions commit.
Thus, the \textsc{RLock()} function (Alg. \ref{alg-TransProcess}, lines 1-13) is provided for transactions to acquire read locks and resolve concurrent read-write conflicts. First, it checks if the transaction already owns the lock as a writer or is listed as a reader. If so, the lock is immediately granted. Otherwise, it checks if another writer exists and applies the Wound-Wait protocol by comparing the priorities: waiting for a higher-priority transaction to commit and aborting the lower-priority transaction.
After acquiring the lock, the transaction is added to the reader list, the read lock counter is incremented, and the row is inserted into the transaction's read set. 

For write operations, the \textsc{WLock()} function (lines 14-34) is used to resolve both write-read and write-write conflicts. Similarly to handling read-write conflicts, the Wound-Wait protocol is applied if conflicts occur. Once the lock is acquired, the row is added to the read and write sets, and the modification is deferred to the commit phase by using the delayed write apply mechanism.

Note that the priority comparison between transactions in \oursys is determined based on the combination of `priority' and `tid' (`startTS + wid'). Specifically, we define \( T_1 > T_2 \) if \( T_1.\text{priority} > T_2.\text{priority} \) or \( (T_1.\text{priority} = T_2.\text{priority} \land T_1.\text{tid} < T_2.\text{tid}) \).
A smaller `tid' indicates that the transaction was initiated earlier. It is important to note that the case where \( T_1.\text{tid} = T_2.\text{tid} \) cannot occur, as `tid' incorporates the worker ID (`wid`) of the thread associated with the transaction, ensuring its uniqueness at the time of assignment.

\subsection{Model Transition and Dynamic Adjustment}
\label{switch_validation}
In \oursys, transactions periodically adjust their priority and locking actions by invoking the RL Model to reduce abort rates. 

\begin{figure}[t]
    \centerline{\includegraphics[width=0.38\textwidth]{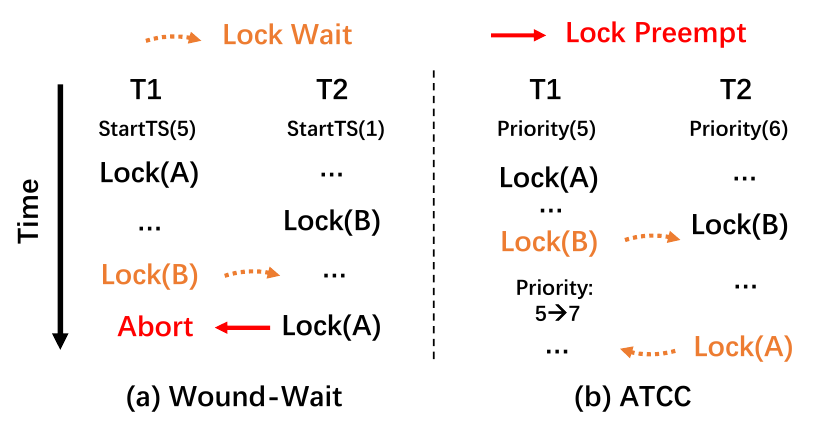}}
    \vspace{-0.1in}
    \caption{Deadlock caused by transaction priority changes when using the Wound-Wait algorithm.}
    \Description[]{}
    \label{fig-DeadLock}
    \vspace{-0.1in}
\end{figure}

\Paragraph{Priority Change and Deadlock Prevention}
\label{deadlock}
The dynamic priority adjustment mechanism reduces abort rates for high-cost or critical transactions, but it violates the assumption of static priorities (e.g., `startTS') underlying traditional deadlock avoidance mechanisms, thereby reintroducing the potential for deadlocks. In Wound–Wait, for instance, each transaction is assigned a static priority (based on `startTS'), and a partial order relation (`< ') is used to break wait cycles. As depicted in Figure \ref{fig-DeadLock}a, transaction $T_2$ has a higher priority than $T_1$ since it starts earlier (i.e., $T_2.startTS=1 < T_1.startTS=5$). $T_1$ and $T_2$ first require the lock on items $A$ and $B$. When $T_1$ subsequently requests a lock on $B$, it must wait because the higher-priority transaction $T_2$ already holds $B$. When $T_2$ requests a lock on $A$, it finds that the lower-priority transaction $T_1$ holds the lock and therefore `wounds' $T_1$ by aborting it and taking over the lock.
Since priorities remain fixed throughout execution and Wound-Wait always aborts the lower-priority transaction in a conflict, wait edges are always directed from lower-priority to higher-priority transactions. This guarantees that the wait-for graph is acyclic and hence deadlock-free.

In contrast, when priorities change at runtime, this invariant is violated. As illustrated in Figure \ref{fig-DeadLock}b, consider a scenario where $T_1$ initially has a lower priority than $T_2$. $T_1$ requests a lock on $B$ and is blocked by $T_2$ due to its lower priority. Over time, the priority of $ T_1$ is dynamically increased, for instance, due to prolonged execution. Later, $T_2$ requests a lock on resource $A$, but is now blocked by $T_1$, whose priority has become higher. This results in a circular dependency, leading to a deadlock.

To mitigate this, when a transaction's priority increases, it re-evaluates its pending lock requests to determine whether it can now preempt the lock or not. This re-evaluation mechanism ensures that dynamically adjusted priorities do not result in unresolved circular waits, thereby preventing the formation of deadlocks. 

\Paragraph{Retroactive Validation and Lock Escalation}
When the locking action changes, the transaction validates the data items that were previously accessed optimistically to detect existing conflicts by invoking the \textsc{ActionChange} function (Alg. \ref{alg-validation}, lines 1-9), ensuring isolation guarantee (e.g., serializability). The transaction first verifies whether the reading data has been modified and attempts to acquire read locks on these data items to prevent further modifications by other transactions. 
Subsequently, the transaction proceeds to resolve write-write conflicts and employ write locks on the data items in the targeted write set. If any read or write lock acquisition fails, this indicates that a conflict with a higher-priority transaction exists. In such cases, the current transaction must abort to prioritize the execution of the higher-priority transaction.

\begin{algorithm}[t]
\begin{spacing}{1}
\SetAlgoNoLine
\small
    \caption{Functions of Validation}
    \Description[]{}
    \label{alg-validation}
    \KwIn{thread (transaction) context $ctx$}
    \KwOut{return true if success; false otherwise}

    \SetKwFunction{FSwitchValidation}{ActionChange}
    \SetKwFunction{FValidation}{Validation}
    
    \SetKwProg{Fn}{Function}{:}{\KwRet}
    \Fn{\FSwitchValidation{LockAction $ac$}}{
        \textcolor{gray}{ //Lock the HRS or RS, HWS or WS based on the lock action.}\\
        \ForEach{$row$, $read\_ver$ in $txn.HRS/RS$}{
            \uIf{$read\_ver$ == $row.curr\_ver$} {
                RLock($row$); \textcolor{gray}{ //Lock the row or abort.}\\
            }
            \uElse {
                Abort($ctx$); \textcolor{gray}{ The row is updated by other txn, abort.}\\
            }
        }
        \ForEach{$row$ in $txn.HWS/WS$}{
            WLock($row$); \textcolor{gray}{ //Lock the row or abort.}\\
        }
    }

    \SetKwProg{Fn}{Function}{:}{\KwRet}
    \Fn{\FValidation{}}{
        \ForEach{$row$ in $txn.WS$}{
            WLock($row$); \textcolor{gray}{ //Lock the row or abort.}\\
        }
        \ForEach{$row$, $read\_ver$ in $txn.RS$}{
            \uIf{$reader\_list$.Find($ctx$.tid)} {
                continue; \textcolor{gray}{ The row is protected by the RLock.}\\
            }
            \While {IsCommitting($row$)} {
                Abort($ctx$); \textcolor{gray}{ //The row is updating by other txn, abort.}\\
            }
            \uIf{$read\_ver$ $\neq$ $row.curr\_ver$} {
                Abort($ctx$); \textcolor{gray}{ The row is updated by other txn, abort.}\\
            }
        }
        \ForEach{$row$ in $txn.WS$}{
            SetCommitting($row$);\\
        }
    }
    
\end{spacing}
\end{algorithm}



\subsection{Unified Commitment}
\label{validation}

Upon committing, \oursys employs a validation procedure (Alg. \ref{alg-validation}, lines 10-21) similar to that of Silo \cite{silo} for all transactions, as some operations are processed optimistically.

The validation begins by attempting to lock all data items in the write set. Unlike Silo, which relies on a single updating latch to lock data items, \oursys leverages WLock to detect potential write-write or read-write conflicts between concurrent transactions with varying priorities. 
After successfully locking the write set, \oursys proceeds to validate the read set. For data items that the transaction has previously accessed under the RLock, validation can be skipped, as the \texttt{read\_ver} and the current row version (\texttt{curr\_ver}) are guaranteed to be consistent. 
For optimistically executed read operations, \oursys performs lightweight version validation instead of acquiring the RLock (lines 16-19). This is because locking these items for validation could introduce unnecessary blocking or premature aborts in the presence of transient read-write conflicts. Since \oursys adopts a \textit{write‑delay apply} mechanism, where modifications are only applied at the final commit stage, such read-write conflicts do not compromise serializability as long as the data has not been updated (e.g., as illustrated by $T_2$ in Figure 4c).

After acquiring the write locks and validating the read set, the transaction marks the write locks as committing, ensuring that optimistic reads cannot access inconsistent data and that higher-priority transactions cannot preempt the acquired locks.
Separating the WLock() operation from SetCommitting() is critical, as the transaction might be blocked during lock acquisition.  Coupling these operations could result in some data items being prematurely marked as is\_committing, but the transaction will not update those data items immediately, introducing unnecessary blocking to other optimistic reads.

After validation, the transaction obtains a commit sequence number (data version number), writes the updates to the Write-Ahead Log (WAL), and applies the modifications to the global rows.


\subsection{Priority-Driven Lock Handover}

After a transaction commits or aborts, it enters a clean-up phase to safely release locks and transfer access rights to waiting transactions based on their priorities.
For each data item in the write set, \oursys first verifies if the current transaction remains the lock holder. If it does, \oursys clears the committing flag, releases the lock, and scans the associated $lock\_queue$ to determine the next eligible transaction(s) to acquire the lock(s). If not (e.g., preempted by a higher-priority writer), the data item is skipped. 
For each row in the transaction’s read set, the system checks if the transaction identifier is still present in the row’s $reader\_list$ and removes its entry from the list, and decrements the read counter. This ensures that shared read locks are fully released, allowing subsequent writers to acquire exclusive access safely. When $R\_count$ reaches zero, the system examines the $lock\_queue$ and assigns the lock to waiting lower-priority transactions as appropriate.

\subsection{Discussion}
In \oursys, an action change never involves releasing locks. Only agentic transactions with a high abort risk and large retry cost choose to acquire locks. Even when conflicts become rare, releasing these locks would expose the previously read data to modifications, and any resulting aborts would incur an unacceptable retry cost. Moreover, most transactions can rely on \textit{delay write apply} to read data optimistically without being blocked by locks. Only those agentic transactions that actually encounter conflicts need to acquire locks to ensure conflict-free execution. Therefore, \oursys does not adopt the release lock action. Once a transaction has escalated to a locking action, it retains that action until it is either committed or aborted.




\section{Evaluation}
\label{section:6}
\label{section:Evaluation}
In this section, we evaluate the performance of \oursys with several agentic workloads.

\Paragraph{Implementation}
\oursys is implemented based on openGauss-MOT \cite{openGauss-MOT}. openGauss is an open-source industrial-grade database system, and openGauss-MOT is an in-memory transactional storage engine designed for high-performance OLTP workloads. Unlike most academic in-memory database prototypes~\cite{Polyjuice, Polaris, Plor, mocc, yu2016tictoc}, which execute transactions in a stored-procedure mode with fixed access patterns, \oursys leverages the SQL interface provided by openGauss-MOT to support flexible query processing. This flexibility allows \oursys to seamlessly integrate with agentic transaction scenarios, which contain various query types, enabling adaptive concurrency control in real-world operational environments.

\Paragraph{Testbeds}
Our experiments run on a server equipped with Intel\textsuperscript{\textregistered} Xeon\textsuperscript{\textregistered} Platinum 8360H CPUs (each with 24 physical cores, 33 MiB LLC, clocked at 3.0 GHz), 768 GiB DDR4 DRAM, and a 2TB NVMe-SSD.  
The operating system is CentOS 7.9 (Linux release 7.9.2009, kernel 3.10). 
To ensure stable performance results, all worker threads are pinned to a single NUMA node, thereby eliminating cross‑NUMA memory access overhead.

\Paragraph{Competitors} In our experiments, we compare \oursys with several representative CC schemes to evaluate the performance:

\begin{itemize}[leftmargin=*]
\item Wound-Wait: A variant of 2PL where a transaction preempts the lock if it is older than the lock holder and waits if younger.

\item Silo \cite{silo}: A representative implementation of OCC and serves as a baseline for high-performance in-memory OLTP systems.

\item MOCC \cite{mocc}: A hybrid CC scheme that selectively acquires pessimistic read locks on hot records during execution and enforces a canonical lock-ordering protocol that allows transactions to acquire, release, and re-acquire locks without deadlocks.

\item Plor \cite{Plor}: A hybrid CC protocol that transactions acquire locks before updating, but do not block on reads. Conflicting reads are performed optimistically and validated in the commit phase.

\item Polaris \cite{Polaris}: An OCC variant that supports transaction priorities via reservations. Each record can be reserved by higher-priority transactions (e.g., repeatedly aborted ones), preventing lower-priority writers from overwriting hot data, reducing tail latency.

\item Polyjuice \cite{Polyjuice}: A learning-based CC framework that models each pre-known transaction operation and the potential conflicts as an execution state, learns a mapping from each execution state to a concurrency control decision table to guide runtime transaction execution with the most suitable concurrency control actions (e.g., whether to wait, which version to read).




  
\end{itemize}

\Paragraph{Workloads} Experiments use three benchmarks, including YCSB \cite{YCSB}, TPC-C \cite{tpcc} and a simulated ticket reservation workload. In YCSB, each query accesses a single record, and the contention level is controlled via a Zipfian distribution. To make it a transactional benchmark, we wrap 10 operations within a transaction and use one table with 10 columns and 1,000,000 rows. 
TPC-C is an industry-standard benchmark for evaluating online transaction processing (OLTP) databases. Only \textit{Payment} and \textit{NewOrder} are used in our evaluation, since several baselines only support these types. Inspired by the characteristics of agentic workloads, we adapt YCSB and TPC-C to emulate agent-like interaction patterns, and refer to the resulting benchmarks as \emph{Agentic-like YCSB} and \emph{Agentic-like TPC-C}. Clients issue agentic transactions with random inter-operation delays (uniformly distributed in 1-20~ms) to model the agent’s extra processing. To model abort handling, when an agentic transaction aborts, we delay its retry by an additional random reasoning time (500~ms-5~ms) to reflect plan reconsideration before re-execution.



The simulated ticket-booking workload represents an agentic Flight-Booking scenario. In this workload, we implement an \textit{agent} using a lightweight \textit{proxy} that communicates with the database through SQL interfaces and connects to \texttt{ChatGPT-4.0} to acquire reasoning and natural language understanding capabilities. The agent operates on a relational schema (including \texttt{Routes}, \texttt{Prices}, \texttt{Aircraft}, and \texttt{Seats}) via a multi-turn ReAct workflow.

\textbf{An Example:} For a travel request (e.g., book SFO$\rightarrow$JFK next Monday, prefer Star Alliance, avoid redeyes, budget $<\$600$), the agent parses the request and generates an initial plan.
Then, it queries the “Routes” table to retrieve flight segments and feasible itineraries. Based on intermediate results, it synthesizes additional SQL queries dynamically to address user constraints, such as filtering by alliance, time windows, or budget by querying “Routes,” “Aircraft,” and “Prices” tables. The agent iteratively optimizes candidate routes, relaxing constraints or exploring alternative joins if needed. Once feasible options are identified, it queries the “Seats” table to verify availability, selects a seat, and executes a write operation to finalize the reservation.

Specifically, in our testing, 80\% of clients issue agentic transactions, while the remaining 20\% continuously execute stored-procedure transactions in the background to represent conventional workloads. To prevent frequent aborts due to accessing hot records, stored-procedure transactions retry after a shorter backoff (10-30~ms). Unless otherwise noted, we report end-to-end transaction latency measured from the initial submission until the transaction eventually commits (including all retries and injected delays). 

To quantify the overhead of LLM interactions, we estimate the average token consumption per transaction, denoted as $T_{avg}$.
Based on profiling the Flight Booking Agent Workload, we set the average tokens per SQL operation $\omega \approx 2703$.
$T_{avg}$ is calculated as $T_{avg} = (1 + R_{abort}) \cdot N_{ops} \cdot \omega$, where $R_{abort}$ is the abort rate and $N_{ops}$ is the number of operations per transaction.

\subsection{Agentic-like YCSB Results}
We first analyze the performance of all the concurrency control algorithms on the YCSB workload with three different variations:

\begin{itemize}[leftmargin=*]
  \item Low Contention: 95\% reads - 5\% writes with a uniform access distribution ($\theta$ = 0.0).

  \item Medium Contention: 90\% reads - 10\% writes with a hotspot of 10\% tuples that are accessed by $\sim$ 50\% of all queries ($\theta$ = 0.7).

  \item High Contention: 50\% reads - 50\% writes with a hotspot of 10\% tuples that are accessed by $\sim$ 75\% of all queries ($\theta$ = 0.99).
  
\end{itemize}

\begin{figure}[t]
  \centering
  \includegraphics[width=0.45\textwidth]{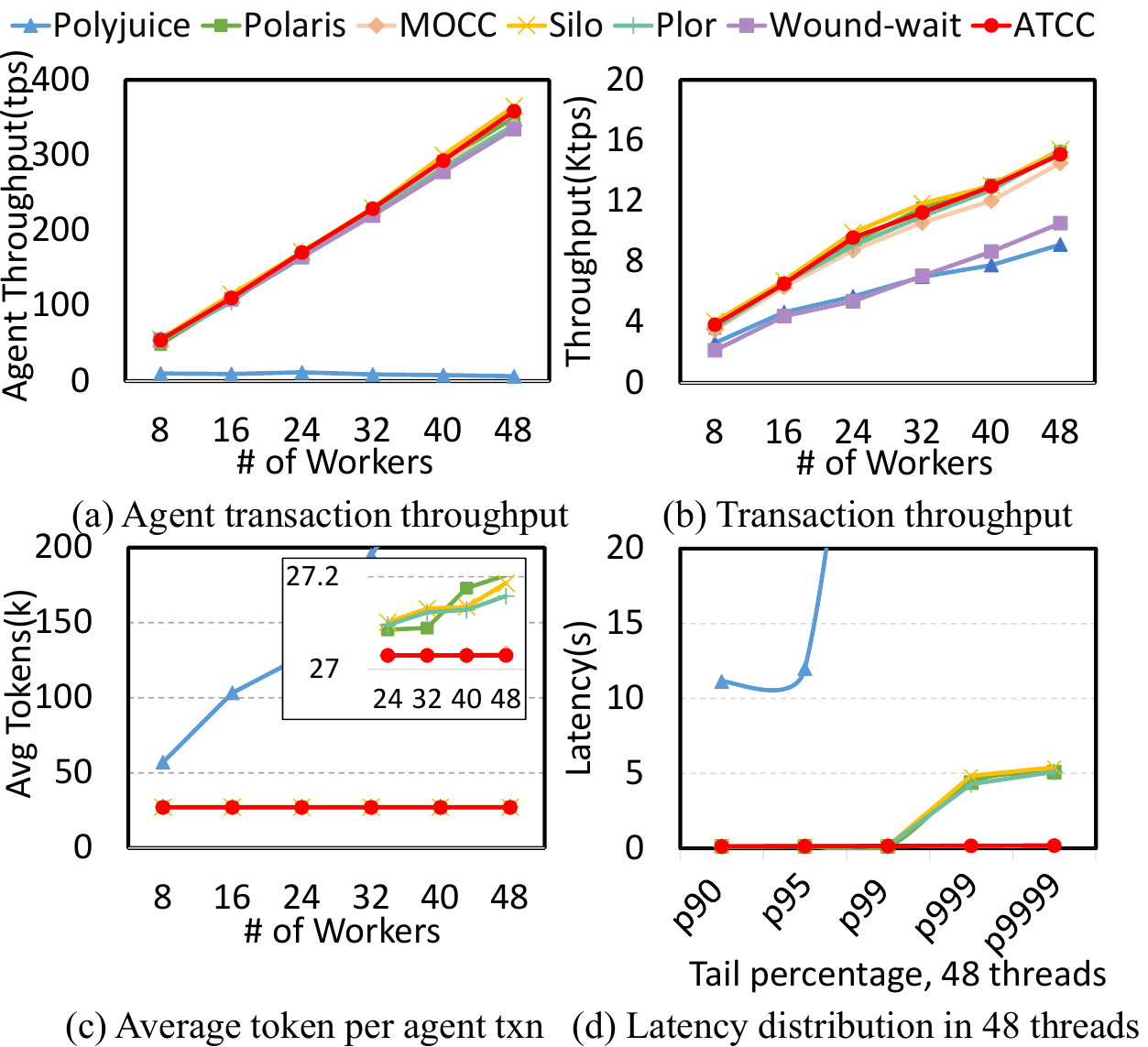}
  \caption{Results when varying the number of workers (YCSB Low-Contention).} 
  \label{fig:ycsb_low}
\end{figure}


\Paragraph{Low contention} 
We first establish a performance baseline using the \textit{Agentic-like YCSB} workload with low contention. 
In this workload, transaction conflicts are rare; thus, system throughput is primarily determined by the inherent implementation overhead rather than conflict resolution efficiency.

As shown in Figures~\ref{fig:ycsb_low}(a) and (b), all concurrency control (CC) protocols scale almost linearly up to 48 threads.
\textbf{\oursys} achieves throughput comparable to Silo.
By performing asynchronous model invocations outside the transaction's critical path and utilizing lightweight latch-free structures for statistics, \oursys minimizes the runtime impact, showing only a margin($\approx 3\%$) compared to Silo’s minimal validation logic.
Polaris, Plor, and MOCC exhibit slightly lower throughput.
Polaris incurs fixed costs for priority reservations and updates on every record access and offers little benefit in conflict-free environments.
Similarly, MOCC introduces approximately 5\% overhead due to its page-level temperature tracking and selective locking mechanisms.
Plor, despite its latch-free design, the bookkeeping overhead of its hybrid pessimistic/optimistic read mode still slightly reduces efficiency in low-contention scenarios.
The learning-based Polyjuice performs the worst in this scenario, as its offline-trained policy, dependent on static transaction types, fails to adapt to the stochastic operations and dynamic access patterns of Agentic-like YCSB, leading to suboptimal decisions and severe performance degradation. 


Figures~\ref{fig:ycsb_low}(c) and (d) report average token costs and tail latencies, reflecting cost efficiency and service quality for agent-centric applications. Most systems maintain low costs and stable latencies.
\oursys maintains a flat token cost ($\approx 27k$), indicating a near-zero abort rate. 
While baselines like Silo and Polaris exhibit slight increases in both token usage and P99.99 latency due to occasional unnecessary retries.

\begin{figure}[t]
  \centering
  \includegraphics[width=0.45\textwidth]{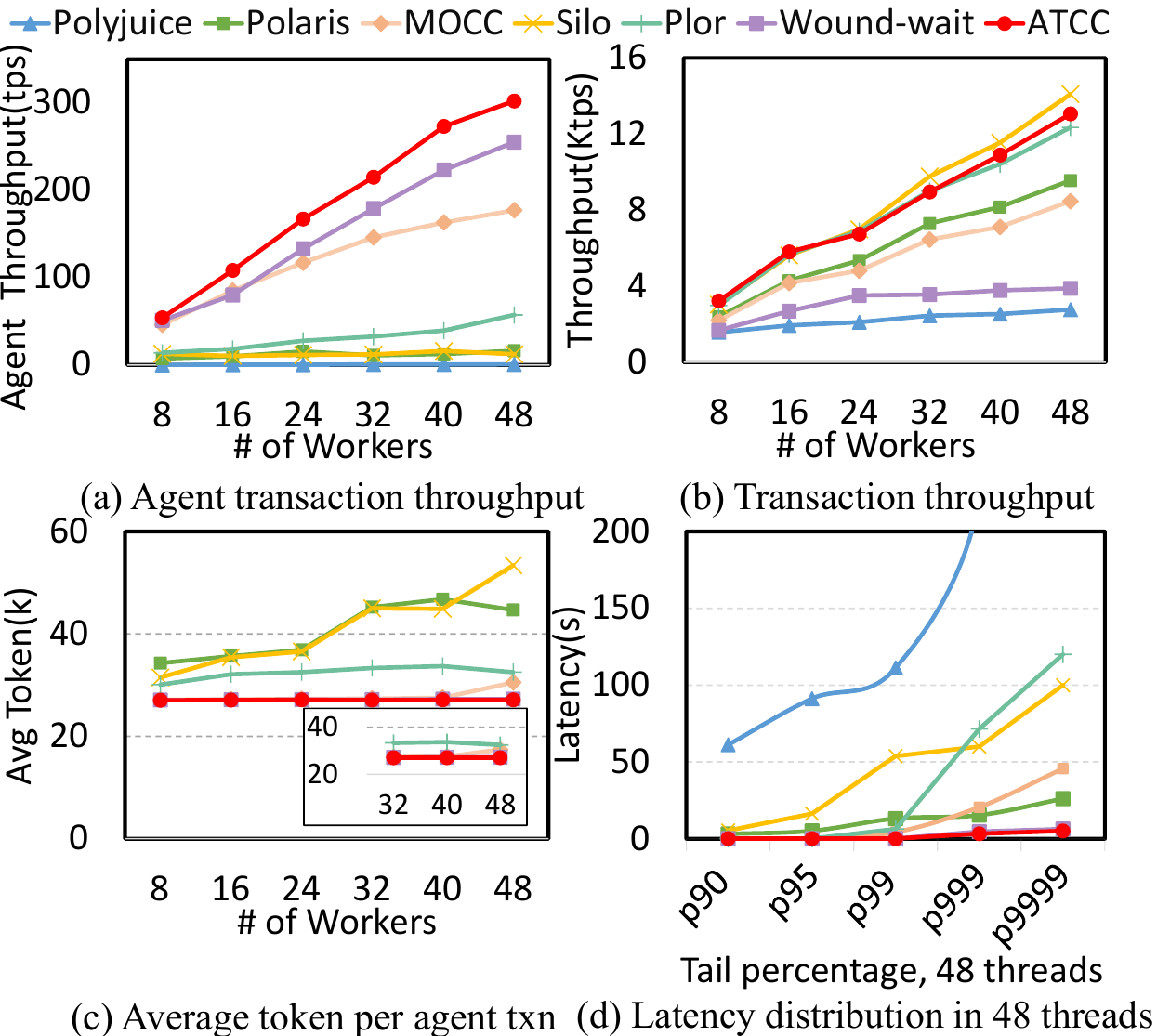}
  \caption{Results when varying the number of workers (YCSB Medium-Contention).} 
  \label{fig:ycsb_medium}
\end{figure}

\Paragraph{Medium contention} 
When contention increases, however, the ways these algorithms handle conflicts begin to significantly affect the performance of long-lived agentic transactions.

As shown in Figure~\ref{fig:ycsb_medium}, all systems continue to scale with increasing worker threads under medium contention, but the performance of agentic transactions degrades to varying degrees depending on each CC scheme's conflict handling and ability to prioritize agentic workloads. 
Silo and Polaris, relying on optimistic validation, suffer from frequent aborts as agentic transactions repeatedly read stale versions of hot keys, as they are repeatedly invalidated by faster stored procedures. 
While the total throughput remains high, this is a misleading metric achieved by cannibalizing agents, leading to the retry loop, and causing agent throughput to collapse (Figure~\ref{fig:ycsb_medium}a).
Wound-Wait improves agent throughput via preemptive locking but degrades the overall system performance due to the indiscriminate blocking of short background transactions. 
MOCC and Plor fall between these two extremes but fail to fully resolve the trade-off: MOCC releases locks prematurely, exposing previously read data to concurrent updates, leading to aborts. Plor avoids concurrent updates using write locks, but in read-intensive workloads, many agentic transactions are still aborted by shorter transactions. 
Polyjuice, optimized for static stored procedures, struggles to adapt to the random access, further impacting performance.


\oursys achieves high throughput by adaptively switching hot data access to pessimistic mode. 
Crucially, by leveraging \textit{deferred updates}, agentic read locks do not block the validation of concurrent stored procedures. 
This strategy achieves a high agentic throughput at 48 threads, higher than Polaris by $18.1\times$ and Polyjuice by $76.9\times$, while reducing token costs by $\sim16\%$ compared to Plor.
Additionally, \oursys remains stable with tail latency, lowering by $5\times$ compared to Polaris.




\begin{figure}[t]
  \centering
  \includegraphics[width=0.45\textwidth]{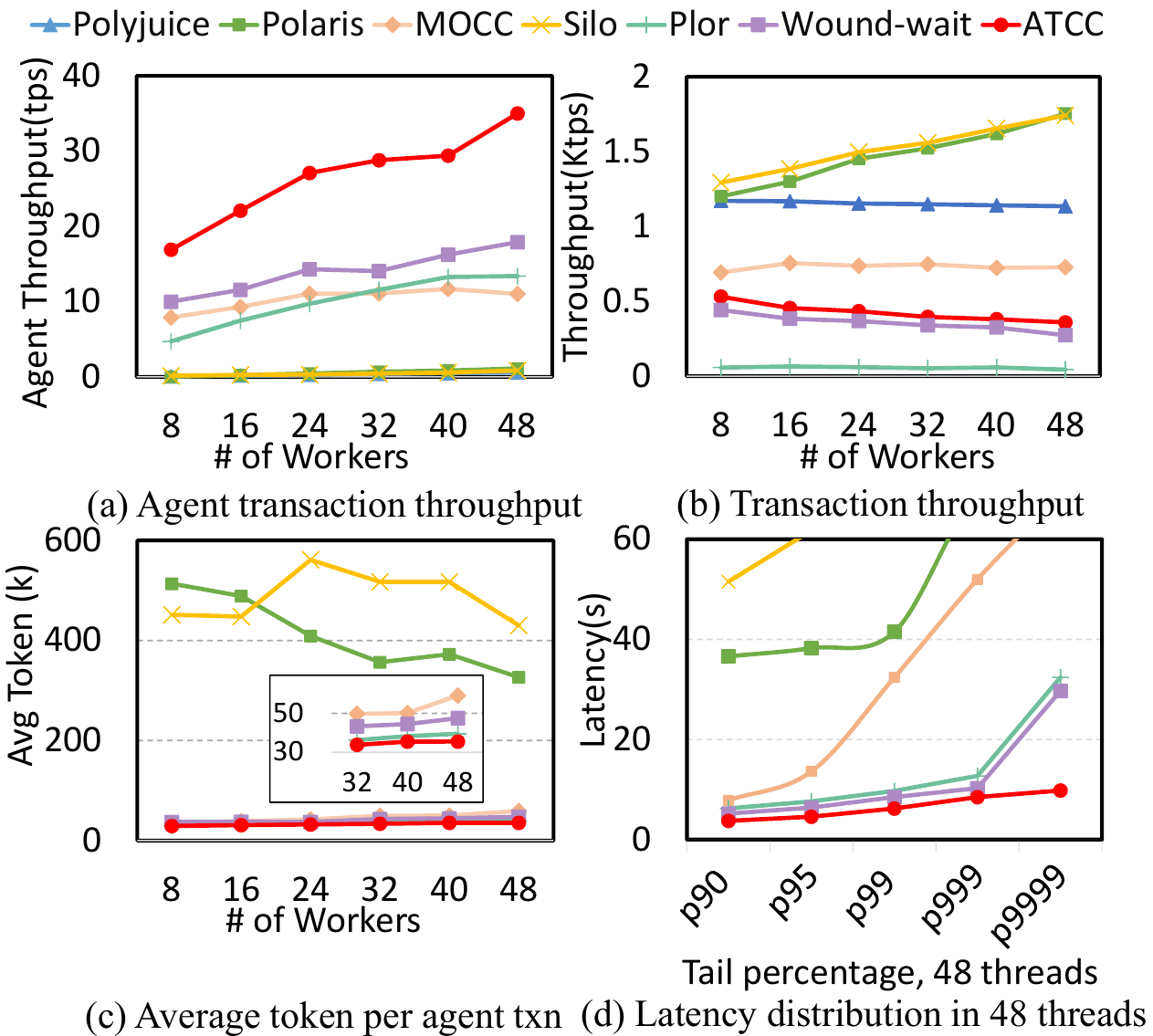}
  \caption{Results when varying the number of workers (YCSB High-Contention).}
  \label{fig:ycsb_high}
\end{figure}


\Paragraph{High contention} 
Figure \ref{fig:ycsb_high} presents throughput over a spectrum of thread numbers in a high contention workload. 
When contention becomes high, hot data items are accessed by many transactions concurrently. Silo, Polaris, and Polyjuice maintain high \textit{total} throughput but suppress agentic transactions to near-zero levels. 
Figure \ref{fig:ycsb_high}(a) highlights \oursys as the only robust solution, outperforming MOCC and Wound-wait by approximately $3.1\times$ and $2.2\times$ in agentic transaction throughput, respectively. While other baselines drop to near-zero throughput due to the retry loops.
This performance advantage stems from \oursys's proactive locking strategy on hot records. Although this protective mechanism reduces short transaction throughput compared with purely optimistic schemes, it ensures fairness and economic viability.


Crucially, \oursys keeps the abort rate below 30\%, reducing token costs by over $15\times$ compared to Silo/Polaris ($\approx 35k$ vs. $500k+$). \oursys exhibits superior tail latency control with fewer retries.
Competitors suffer from unbounded latency growth due to uncontrolled retries, \oursys commits the vast majority of agentic transactions with bounded delay, lowering the P99.99 latency by approximately $3\times$ compared to Wound-wait.


\subsection{Agentic-like TPC-C Results}

\begin{figure}[t]
  \centering
  \includegraphics[width=0.45\textwidth]{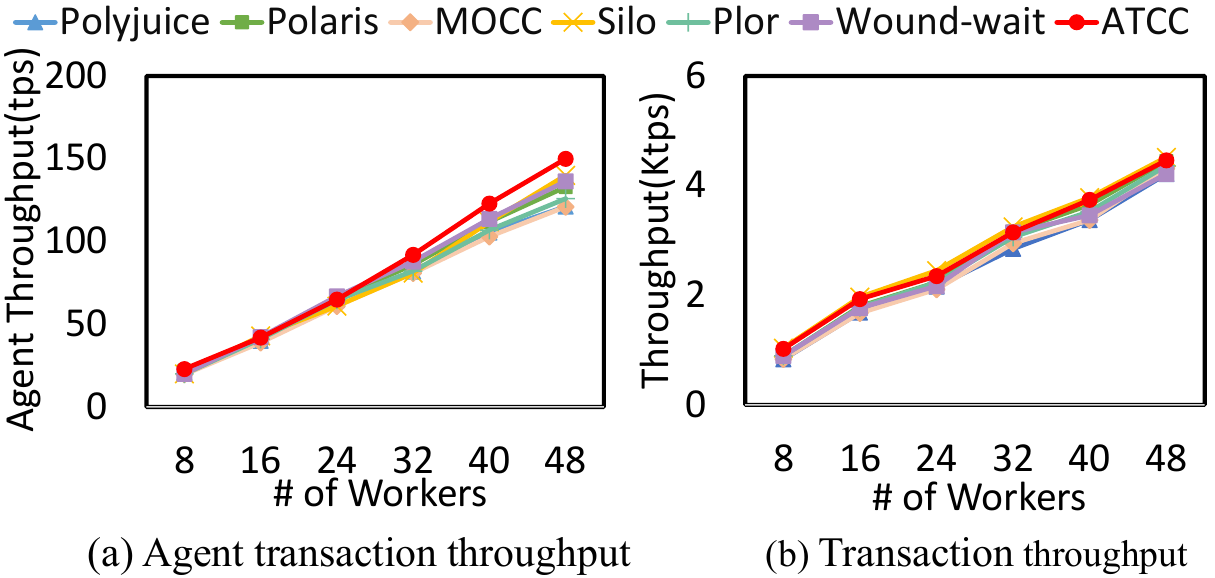}
  \caption{Results for the different CC algorithms (TPC-C with 100 warehouse). } 
  \label{fig:tpcc_low}
\end{figure}

\begin{figure}[t]
  \centering
  \vspace{-10pt}
  \includegraphics[width=0.45\textwidth]{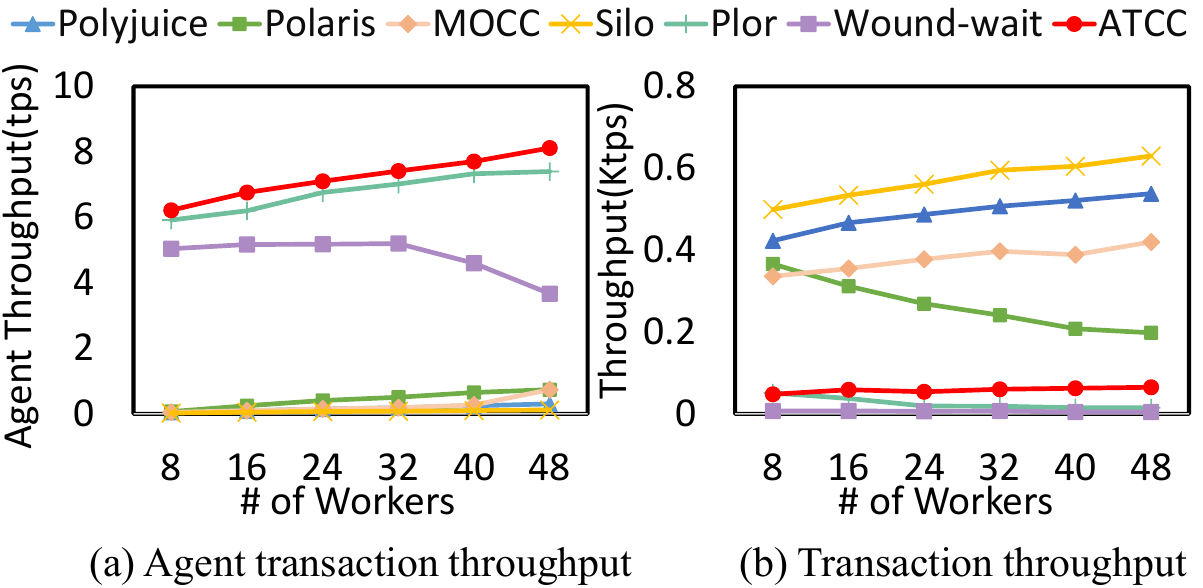}
  \vspace{-0.1in}
  \caption{Results for the different CC algorithms (TPC-C with 1 warehouse).} 
  \vspace{-10pt}
  \label{fig:tpcc_high}
\end{figure}

We now evaluate the concurrency control protocols using the Agentic-like TPC-C workload under two different contention levels by using 1 and 100 warehouses. 
The number of warehouses in TPC-C determines both the size of the database and the amount of concurrency. The warehouse is the root entity for almost all of the tables in the database. 
We first run TPC-C with 100/48 warehouses to simulate a low-contention environment rich in parallelism and then with 1 warehouse to represent high contention where transactions compete heavily for the same set of records.

\Paragraph{Low Contention}
Figure \ref{fig:tpcc_low} shows the throughput results for the 100-warehouse case.
When contention is low, all protocols scale almost linearly with increasing worker threads, consistent with the YCSB-Low-Contention results. Unlike YCSB’s random accesses, each TPC-C transaction has a fixed sequence of operations and stable data dependencies. This determinism allows \textbf{Polyjuice}, which uses an offline pre-trained policy, to allocate nearly optimal concurrency-control strategies for each operation, thus exhibiting excellent scalability similar to Silo.

\Paragraph{High Contention}
Figure \ref{fig:tpcc_high} presents the results for the 1-warehouse configuration, where contention is high due to intense competition for the same records. 
Silo and Polaris suffer sharp throughput degradation of agentic transactions due to validated aborts.
Plor exhibits the lowest performance, bottlenecked by both write blocking and read-conflict invalidations.
MOCC and Polyjuice achieve relative stability using explicit hot-term locking and deterministic policy learning, respectively, but they lack specific mechanisms to prevent long-running proxies from running out of resources.


In contrast, \oursys ensures the viability of agentic transactions by actively protecting their read and write sets.
By enforcing pessimistic isolation on hot records, \oursys achieves an agentic throughput $2.2\times$ higher than Wound-Wait and over $10\times$ higher than other baselines.
This confirms that under pathological contention, targeted protection is the only viable strategy to preserve interactivity stability.


\subsection{Agentic Workload Performance}

\begin{figure}[t]
  \centering
  \includegraphics[width=0.45\textwidth]{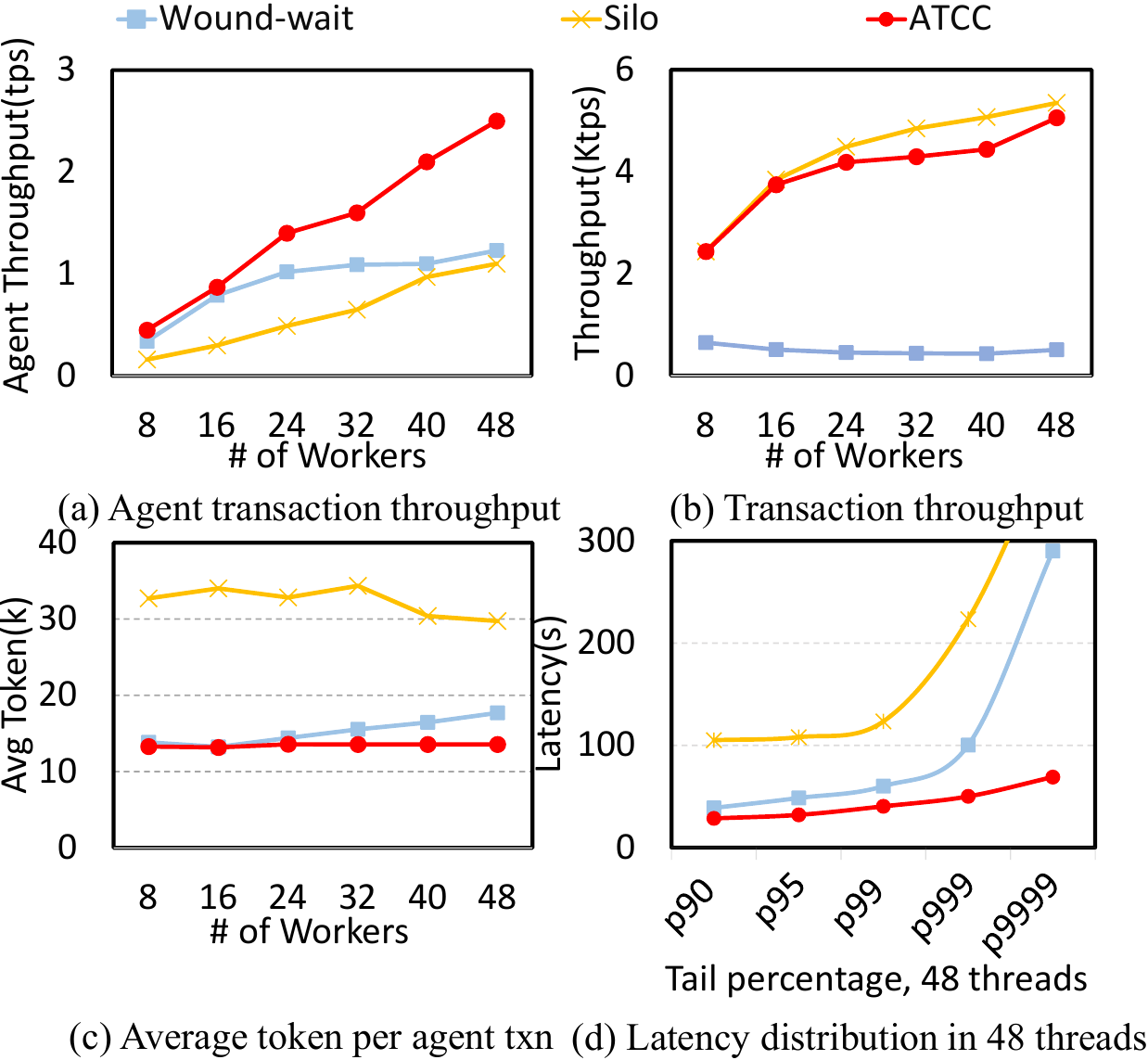}
  \caption{Performance comparison on the Flight Booking agentic workload.}
  \label{fig:agent_tps}
\end{figure}

In this experiment, we evaluate how \oursys performs on an agentic ticket-ordering workload.
Since MOCC, Plor, Polaris, and Polyjuice do not support standard SQL interfaces, we just compare \oursys with Wound-Wait and Silo under the same code-base. Figure~\ref{fig:agent_tps} presents the performance results under the Flight Booking workload (80\% agentic mixed).

As shown in Figure~\ref{fig:agent_tps}(a), \oursys achieves a peak agent throughput of $\sim$2.5 tps, outperforming Wound-Wait by $2\times$ and Silo by $2.5\times$.
Crucially, this gain does not compromise background tasks.  
In Figure~\ref{fig:agent_tps}(b), \oursys maintains a total throughput ($\sim$5k tps) comparable to Silo. 
Employing adaptive locking with priority-aware scheduling, ATCC effectively protects long-running agents from starvation without inducing system-wide gridlock. 
Conversely, Silo achieves high total throughput only by starving agentic transactions, rendering it practically unusable for agent-centric applications. Similarly, Wound-Wait's indiscriminate locking triggers severe contention and preemptive aborts of agentic transactions, causing total throughput decreased by nearly $10\times$ compared to \oursys.

Figure~\ref{fig:agent_tps}(c) and (d) highlight the economic implications and tail latency. 
Silo incurs prohibitive costs ($\sim$33k tokens) due to pathological retry loops, in which agents repeatedly waste tokens on reasoning only to fail validation. \oursys eliminates this waste via adaptive locking, reducing token costs by approximately $60\%$ ($\sim$13k).
Regarding tail latency, \oursys maintains a bounded P99.99 latency of $\sim$70s.
This represents a $4\times$ reduction compared to Wound-Wait ($\sim$290s), which suffers from significant lock queuing delays and aborts under high concurrency.

\subsection{Variable Agentic Client Proportion}

\begin{figure}[t]
  \centering
  \includegraphics[width=0.45\textwidth]{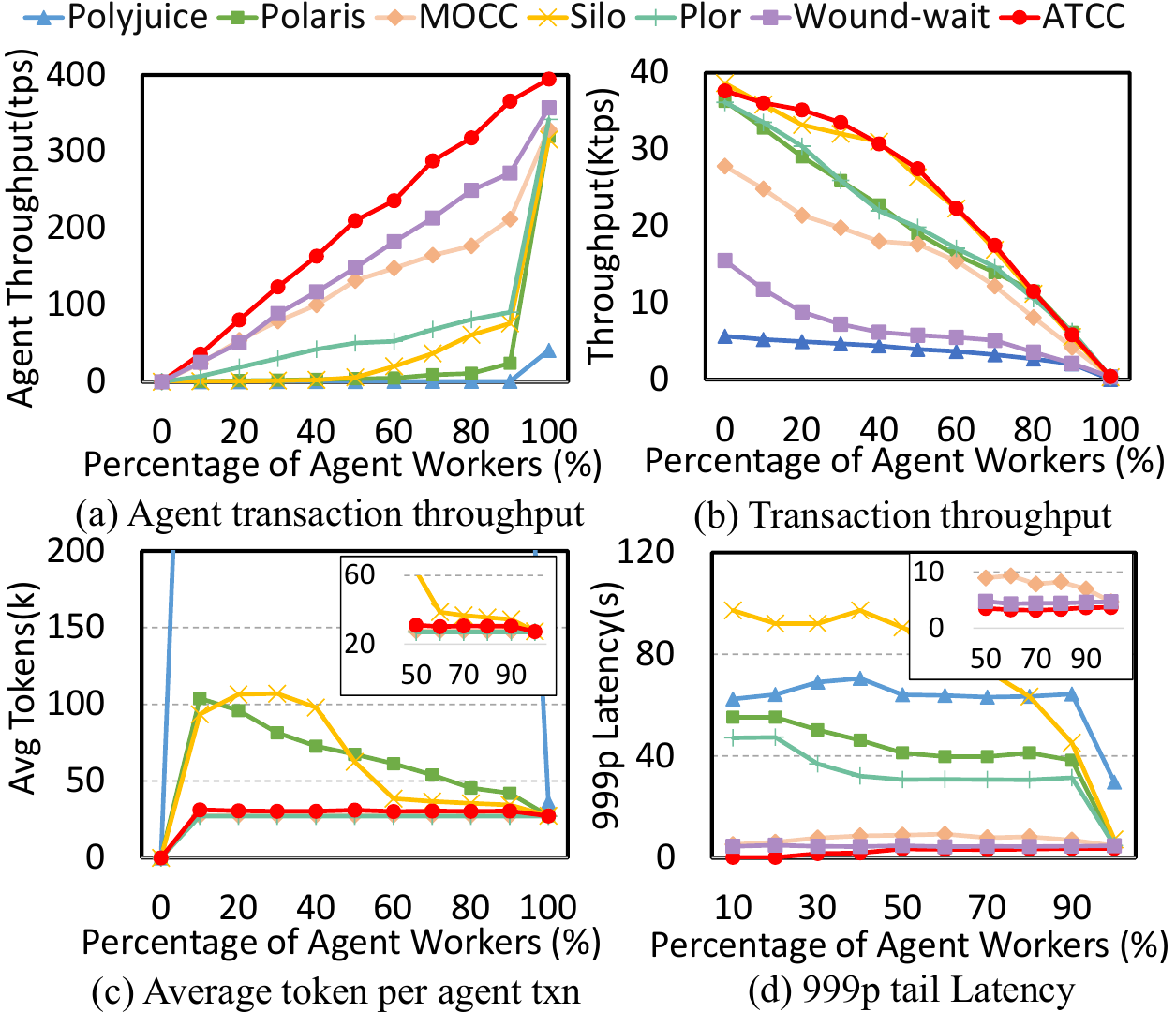}
  \caption{Performance with varying interactive client proportions (YCSB Medium-Contention).} 
  \label{fig:vary_client}
\end{figure}


Real-world deployments exhibit dynamic shifts between interactive and background-heavy phases. A practical CC scheme must therefore remain robust as the mix of agentic and stored‑procedure clients shifts over time.
To evaluate robustness under such fluctuations, we vary the proportion of agentic workers from 0\% to 100\% (48 threads total) using the YCSB medium contention workload.

As illustrated in Figure~\ref{fig:vary_client}, while total throughput naturally declines as long-running transactions replace short procedures, \oursys demonstrates exceptional robustness.
It achieves nearly linear growth in agentic throughput, maintaining a negligible abort rate ($\sim 0.2\%$) across all mix ratios. By dynamically switching hot agentic operations to lock-protected mode while keeping background tasks optimistic, \oursys maintains minimal token costs ($\sim$27k) and prevents latency spikes.

In contrast, baselines employing uniform conflict resolution fail to protect agent transactions. 
When agentic clients are few (10\% - 50\%), long-running agents are repeatedly invalidated by high-frequency background updates.
This inefficiency manifests as a rapid escalation in costs: Silo's token consumption spikes to $\sim$100k - $3.3\times$ higher than \oursys, while Polaris suffers from tail latency spikes nearly $10\times$ worse than \oursys.


\section{Related Work}
\label{section:7}
\label{sec:related}
\label{section:Related Work}

\Paragraph{Learning-Based Transaction Management}
\label{learning-based-tm}
\pzs{Machine learning techniques have been applied to enhance the performance and adaptability of database transaction management. \cite{Scheduling2019} analyzes potential conflicts from SQL predicates and assigns high-conflict transactions to dedicated FIFO queues for serial execution. However, it requires the complete set of SQL statements to be known before execution, making it inapplicable to agentic workloads where queries are generated incrementally by LLMs.}
Polyjuice~\cite{Polyjuice} treats CC as a fine-grained policy that maps per-operation execution states to actions and trains offline to obtain an optimized state-action policy table. While effective for deterministic stored procedures, its static policies fail to adapt to the stochastic, non-deterministic access patterns of interactive agents, leading to policy misalignment and performance degradation.
\oursys employs a reinforcement learning framework explicitly designed for agentic stochasticity. 
Rather than relying on static templates or pre-declared query sets, \oursys analyzes real-time system contention and phase-level access characteristics to adjust locking scopes dynamically.
This adaptive approach ensures both economic efficiency and service quality, addressing the unique availability constraints of agent-centric applications. 

\Paragraph{\textbf{Hybrid Concurrency Control (Hybrid CC)}}
Hybrid CC combines the strengths of OCC and PCC to handle diverse workloads.
Composing CC \cite{ComposingCC} provides a framework for safely composing different CC protocols.
Dynamic OCC \cite{combiningCC} executes transactions optimistically, then re-executes aborted ones using 2PL based on the collected read/write set during the previous run. 
CormCC \cite{CormCC} partitions data based on historical access patterns and assigns each partition a fixed policy (OCC or PCC), dynamically adjusting boundaries to adapt to contention/hotspot shifts.
MOCC \cite{mocc} integrates ordered locking into PCC to reduce deadlocks. Plor \cite{Plor} enhances wound-wait PCC with optimistic reads, reducing unnecessary lock blocking.

However, all of them face challenges in agentic workloads.
Dynamic OCC assumes \textit{deterministic re-execution}, locking the observed read/write set. This fails with non-deterministic LLM agents that may generate entirely different SQL statements upon retry, rendering captured locks useless. 
For CormCC, its coarse-grained, partition-level adaptation ignores the heterogeneity in agents' behaviors accessing the same data, resulting in mismatched strategies. 
MOCC and Plor lack semantic awareness. They treat token-expensive agent transactions the same as traditional transactions, often leading to the preemption of high-value tasks.

\Paragraph{\textbf{Priority-Based Concurrency Control}}
Priority-based concurrency control is a suitable scheme to prioritize certain classes of transactions over others, ensuring low tail latency of critical transactions. It intuitively seems suitable for prioritizing interactive transactions, aligning closely with our approach.
Existing systems such as Microsoft SQL Server \cite{sqlserver}, Oracle Berkeley DB \cite{berkeleydb}, and CockroachDB \cite{taft2020cockroachdb} incorporate ad hoc prioritization mechanisms to favor high-priority transactions in scheduling or validation. 
More structured academic approaches, such as Polaris~\cite{Polaris} and Ding et al.~\cite{DingPriority}, refine this by integrating priority into optimistic validation. 
Polaris uses a lightweight reservation mechanism based on timestamps and abort counts to guarantee progress for older, retry-heavy transactions, while Ding et al. defer conflict resolution to validate prioritized batches sequentially.

However, relying on static heuristics (e.g., timestamps or abort counts) proves insufficient for agentic workloads.
In optimistic settings, over-prioritizing conflicting high-priority transactions triggers aborts among agentic transactions, while in pessimistic modes, aggressive lock preemption leads to starvation.
Crucially, these static schemes lack semantic awareness of \textit{transaction value}. 
They fail to account for the unique runtime characteristics of interactive agents, such as variable urgency, evolving access patterns, and token consumption.
\oursys overcomes this by defining priority not as a static label, but as a dynamic function of real-time contention and accumulated token investment, ensuring that high-value agents survive without crippling system-wide throughput.

\section{Conclusion}
\label{section:8}
\label{sec:conclusion}
In this paper, we present \oursys, an adaptive concurrency control framework tailored for the emerging paradigm of agentic transactions. 
\oursys introduces a reinforcement learning-based policy to dynamically adapt execution strategies, switching between optimistic and pessimistic modes based on the runtime interpretation of agent behaviors. 
Transactions identified as high-risk or cost-sensitive are proactively protected via lock scheduling, while others proceed optimistically to maximize concurrency. 
By effectively balancing the trade-off between immediate blocking overhead and future abort costs, \oursys addresses the unique challenges of long-running, non-deterministic agentic workflows. 
It preserves the high throughput advantage of traditional schemes while providing the robustness and low tail latency required for LLM-driven data agents.



\bibliographystyle{ACM-Reference-Format}
\bibliography{9all}

\end{document}